\documentclass[aps,pre,twocolumn]{revtex4} 

\usepackage{graphicx}
\usepackage{amsmath}

\usepackage[utf8]{inputenc}
\usepackage{amssymb}

\begin{document}

\title{The effect of initial conditions on the electromagnetic radiation generation in type III solar radio bursts}

\author{H. Schmitz\footnote{Now at: Central Laser Facility, Rutherford Appleton Laboratory, Chilton, 
Oxon., OX11 0QX, United Kingdom} and D. Tsiklauri}
\affiliation{School of Physics and Astronomy, Queen Mary University of London, London, E1 4NS, United Kingdom}

\begin{abstract} 
Extensive particle-in-cell simulations of fast electron beams injected in a background
magnetised plasma with a decreasing density profile were carried out. These simulations
were intended to further shed light on a newly proposed mechanism for the generation of
electromagnetic waves in type III solar radio bursts [D. Tsiklauri, Phys.\ Plasmas,
\textbf{18}, 052903 (2011)]. The numerical simulations were carried out using different
density profiles and fast electron distribution functions. It is shown that 
electromagnetic L and R modes are excited by
the transverse current, initially imposed on the system. In the course of the
simulations no further interaction of the electron beam with the background
plasma could be observed.

%\keywords{TNSA \and Sheaths}
%\PACS{Some pacs numbers}
\end{abstract}

\maketitle

\section{Introduction} \label{SecIntroduction}

It is widely accepted that there is a correlation between super-thermal electron
beams and type III solar radio bursts
\cite{1958SvA.....2..653G,1981ApJ...251..364L}. Whilst the correlation is an
established fact, the actual mechanism that generates the type III burst
emission is not yet fully determined. The main source of the uncertainty is
our current inability to send in-situ probes at distances $0.15-1.5 R_{sun}$  from
the solar surface (photosphere). The most widely accepted mechanism, that
historically appeared first is the plasma emission \cite{1958SvA.....2..653G}.
In plasma emission mechanism quasilinear theory, kinetic Fokker-Planck type
equation for describing the dynamics of an electron beam is used, in conjunction
with the spectral energy density evolutionary equations for  Langmuir and
ion-sound waves. Further, non-linear wave-wave interactions between Langmuir,
ion-acoustic and electromagnetic (EM) waves produce emission at electron plasma frequency,
$\omega_{pe}$ or the second harmonic, $2 \omega_{pe}$ \cite{mmcp2005}. A variant
of the plasma emission mechanism is the stochastic growth theory
\cite{1992SoPh..139..147R}, where density irregularities produce a random
growth, in such a way that Langmuir waves are generated stochastically and
quasilinear interactions within the Langmuir clumps cause the beam to fluctuate
about marginal stability. The latter models have been used for producing the
solar  type III burst observable parameters \cite{2008JGRA..11306104L}. Other
possible mechanisms include: linear mode conversion \cite{2005PhPl...12e2315C},
antenna radiation \cite{2012ApJ...755...45M} and non-gyrotropic electron beam
emission \cite{2011PhPl...18e2903T}.

Recent works \cite{2012PhPl...19k0702P,2012PhPl...19k2903P} elucidated further
the non-gyrotropic electron beam emission, first proposed in
Ref.\cite{2011PhPl...18e2903T}. In particular, the effect of electron beam pitch
angle and density gradient on solar type III radio bursts was studied
\cite{2012PhPl...19k2903P} and the role of electron cyclotron maser (ECM)
emission with a possible mode coupling to the z-mode was explored
\cite{2012PhPl...19k0702P}. In this paper, using large-scale Particle-In-Cell (PIC)
simulations, we explore the non-gyrotropic electron beam emission mechanism by
studying the effects of electron beam kinetics and $(\omega,k)$-space drift, in long term
evolution of electromagnetic emission generation of type III solar radio bursts.
The following improvements and progress in understanding of the radio emission
mechanism are made: (i) Improved numerical simulations with larger spatial
domain and longer end-simulation times; (ii) The electron beam injection on a
density plateau followed by a decreasing density gradient the latter mimicing the
Sun-earth system; (iii) Consideration of a ring and shifted ring electron
initial velocity distribution functions; (iv) The role of the $(\omega,k)$-space drift in
the radio emission;  It is worthwhile to note that
Ref.\cite{2012PhPl...19k0702P} proposed mode coupling on the density gradient as
a source of radio emission as opposed to the $(\omega,k)$-space drift advocated in the
present work. The situation is analogous to the auroral waves emitted near the
plasma frequency in Earth auroral ionosphere \cite{2011JGRA..11612328L}.

\section{Simulation Setup \label{SecSimulation}}

We simulate the injection and traversal of the electron beam through a density
gradient using the fully relativistic particle in cell simulation code EPOCH.
The code uses Boris  scheme \cite{Boris:1970,Birdsall:1985} for advancing the
particles together with the charge conserving method by
Esirkepov \cite{Esirkepov:2001} for calcultaing the currents. The shape
function of the particles, and thus the order of the weighting scheme, can be
chosen between top-hat (second order), triangular (third order) and spline
interpolation (5th order). The electromagnetic fields are solved using the
finite difference time domain (FDTD) algorithm on a Yee
grid\cite{Yee:1966,Taflove:2005}. Open boundaries have been implemented using
convolutional perfectly matched layers (CPML) \cite{Roden:2000} and were ported
into EPOCH from a previuos code for simulating electromagnetic pulse propagation
in transparent media\cite{Schmitz:2012a}.

We perform 1-dimensional calculations of the density gradient between Sun and
Earth. The density drops from $n_e=n_S=10^{14}\text{m}^{-3}$ near the Sun to
$n_e=n_E=10^{10}\text{m}^{-3}$ near the Earth. The density is assumed to follow
a parabolic profile
\begin{equation}
n_e(x) = n_E + (n_S-n_E)(x/L-1)^2, \label{EqDensityProfile}
\end{equation}
where $L$ is the size of the simulation box and $x$ is the coordinate along the
1-dimensional simulation domain. In some simulations, the density profile is
extended towards the left by a constant density plateau with the maximum density
$n_S$. The extent of this plateau is chosen to be $L/10$. The background plasma
temperature is assumed to be $T_e = T_i = 10^5$K. Due to computational
restrictions we are limited to 65000 grid points for most of the simulation runs
and to 130000 grid points for two larger runs, hereafter denoted as \emph{long
runs}. In order to avoid numerical heating, PIC simulations usually have to
resolve the Debye length. The minimum of the Debye length in the system is at
the location of the highest density, near the Sun and evaluates to
$\lambda_D=2.18\times10^{-3}$m. Because of these limitations, we cannot hope to
simulate the complete system with a 1:1 scaling. We are therefore forced to
compress the distance between Sun and Earth do a manageable length. We choose a
system length of 
$L=245.7$m for the standard runs and $L=491.4$m for the long runs. This means
that we under-resolve the Debye length by a factor of approximately 2. In order
to suppress numerical heating the 5th order spline interpolation is chosen for
the particle weighting in all runs. While numerical heating will still be
present, it takes place on time-scales much longer than the simulation time. It
can, therefore, be neglected. Performing under-resolved simulations using a
higher order interpolation scheme is frequently employed when studying
interaction of lasers with solid density plasmas \cite{Schmitz:2012b}.

While the box length $L$ is still much shorter than the Sun-Earth distance, we
anticipate that the underlying physics do not change drastically. Naturally,
there are limitations to this. Any effect that increases with the density
gradient will be artificially enhanced. On the other hand we will only be able
to observe effects that happen on very short time scales and long time scale
effects will not be present in the simulations.

In previous investigations\cite{2011PhPl...18e2903T,2012PhPl...19k0702P} the
density profile was chosen to be symmetrical, i.e.\ $L$ in equation
(\ref{EqDensityProfile}) was replaced by $L/2$, and boundary conditions were
periodic. This had the consequence that only half of the simulation domain could
be used. Here we simulate only one half of the symmetric parabolic density
profile. This is achieved by using open (absorbing) boundary conditions on both
sides. In order to avoid rarefaction waves due to the escaping of particles at
the boundaries, the background plasma density tends towards zero in a thin layer
near the boundaries of the simulation domain. 

The beam is set up as a separate electron population near the left boundary of
the domain with a density profile specified by
\begin{equation}
n_b(x) = n_0 \exp\left[-\left(\frac{x-L/25}{L/40.0}\right)^8\right],
\end{equation}
where $n_0=10^{11}\text{m}^{-3}$ is the beam density. This means that the beam
density has a roughly flat profile with a length of $L_{b} = L/20$. When a
density plateau is present the initial beam profile will sit completely within
the constant density region. We perform these runs in order to investigate the
influence of the density gradient at the injection location on the emitted
radiation. The beam temperature is assumed to be $T_b = 2\times10^6$K and the
average speed of the beam electrons is $v_b=0.2c$, where $c$ is the speed of
light. The electron distribution function assumed to be thermal in the
$v_x$--$v_{\perp}$ plane, where $v_x$ is the velocity parallel to the simulation
domain and $v_{\perp}$ is the velocity perpendicular to the simulation domain. 
The drift velocity is assumed to be equal in the $x$ and the perpendicular
directions,
\begin{equation}
f_b(x,v_x,v_{\perp}) = \frac{n(x)}{2\pi m_e kT_b}
\text{e}^{-\frac{(v_x-v_b/\sqrt{2})^2 - (v_{\perp}-v_b/\sqrt{2})^2}{2m_ekT_e}},
\label{EqDistributionPerp}
\end{equation}
where $m_e$ is the electron mass and the distribution function has been
integrated over the azimuthal angle in the perpendicular velocity plane
$v_y$--$v_z$. Due to this integration, the choice of the distribution function
is not unique. We have initialised the electron beam with two different
distribution functions in the perpendicular plane. The first choice represents a
beam, purely in the $v_y$ direction,
\begin{equation}
f_b(x,v_x,v_y, v_z) = \frac{n(x)}{2\pi m_e kT_b}
\text{e}^{-\frac{(v_x-v_b/\sqrt{2})^2 - (v_y-v_b/\sqrt{2})^2}{2m_ekT_e}}.
\end{equation}
In the following we will refer to this distribution as the \emph{directed beam}
distribution. The second choice corresponds to a ring in the $v_y$--$v_z$ plane,
\begin{equation}
f_b(x,v_x,v_y, v_z) = \frac{n(x)}{2\pi m_e kT_b}
\text{e}^{-\frac{(v_x-v_b/\sqrt{2})^2 - (\sqrt{v_y^2 +
v_z^2}-v_b/\sqrt{2})^2}{2m_ekT_e}}.
\end{equation}
This distribution will be called the \emph{ring distribution}. Note that both
distribution functions result in the same distribution function when integrated
over the azimuthal angle in the $v_y$--$v_z$ plane.

In addition, we impose a constant background magnetic field of $B_0=3.2$G along
the $x$-axis. The choice of the magnetic field results in a ratio of cyclotron
frequency to plamsa frequency of $\omega_C / \omega_{pe} = 0.1$, at $x=0$, which means the
plasma is weakly magnetised. The density gradient, together with the constant
magnetic field and constant background temperature, implies that the plasma is
not in pressure balance. This is justified due to the fact that the solar wind
itself is not in pressure balance, as it continually flows outward. Due to the
short times we do not observe any tendency towards a pressure equilibrium during
the course of the simulations.

All three particle species, ions, background electrons and beam electrons are
initialised with 500 particles per cell. Due to the localisation of the beam,
however, the total number of beam particles is substantially less than that of
the background electrons. The simulations were run on 512 cores and took
typically around 40h.

We would like to note that the chosen length scales and parameters are more
suited for studying the EM wave generation in the vicinity of the Sun rather than addressing the 
question radio burst propagation between the Sun and Earth. The reason is 
three-fold: (i) We wish to focus on solar coronal type III radio bursts, not the inter-planetary ones, which have
much longer durations and hence different frequency drift ranges.
(ii) The parameter choice is not vital for the physics of mechanism itself. For example, 
our near-Earth density is taken 10$^{10}$ m$^{-3}$, which is not
realistic, of course, but the exact value of $n_e$ is not important for the result. 
Due to the parabolic shape of the density
profile, there is only a noticeable difference between our density
profile and one with $n_e=10^6$ m$^{-3}$ (the realistic value) in only the rightmost 5\% of the
simulation domain. That is not where any important physics takes place. The distances in our graphical
results are normalised to $c/\omega_{pe}$, thus, in principle, by adjusting the number density, the spatial re-scaling can be
performed to up-scale the results to the real Sun-Earth distance.
The chosen beam density $n_b=10^{11}$ m$^{-3}$ is essentially a free parameter, which only sets the time-scale of quasi-linear relaxation 
(Sec. IV, however, mentions that some models, such as plasma emission,
put their constraints).
The chosen magnetic field, circa 3 Gauss, is realistic for some heights above solar coronal active regions, but the 
main constraint is the need to satisfy Maxwell's equation $div {\vec B}=0$,
in one dimension. Thus we cannot allow magnetic field variation in $x$. Our model also ignores a variation of temperature with distance, but we think its effect is
probably not essential -- the density gradient is, however.
(iii) The aim of this work is to study micro-plasma-kinetic physics of the radio emission {\it generation}. The present results are state-of-the-art what can be achieved with the
current, available to authors', computational resources -- 512 parallel computer processors. The radio burst {\it propagation} between the Sun and Earth
are addressed by another set of models, which use quasi-linear plasma theory that does not resolve micro-physics and hence cannot address the question of the
generation mechanism of the radio emission.

\section{Results}

\begin{figure}
\begin{center}
\includegraphics[width=7cm]{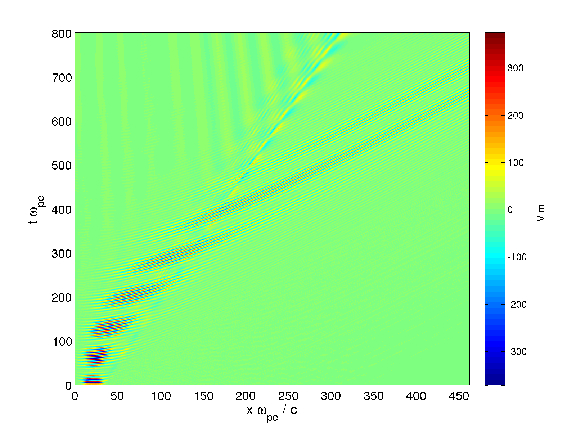}\\
\includegraphics[width=7cm]{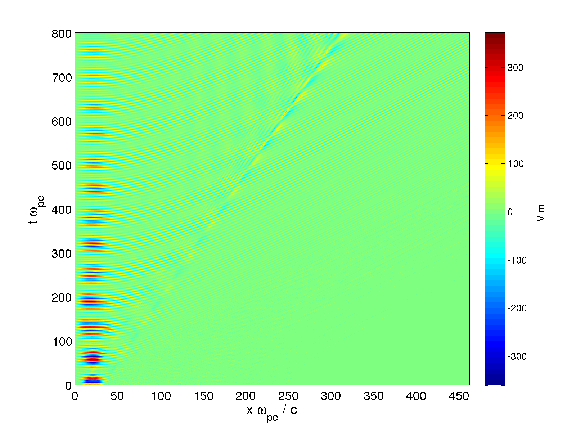}
\end{center}
\caption{The electric field component $E_y$ as a function of space and time for
the reference run (top panel), with a density plateau (lower panel)}
\label{FigEyRefPlateau}
\end{figure}

\subsection{Density Gradient at Injection Location}

We first investigate the influence of the density gradient at the injection
location. Figure \ref{FigEyRefPlateau} shows the perpendicular electric field
component $E_y$ as a function of position $x$ and time $t$. The top panel shows
the result from the reference run in which the Maxwellian electron beam is
injected into the density gradient. The plot clearly shows that the electric
field oscillates roughly with the plasma frequency from the onset of the
simulation. The oscillation is restricted to the beam region but shows no other
spatial dependence initially. This indicates that the oscillation is imposed by
the initial conditions linked to the strong $j_y$ current present in the initial
distribution function. In addition to the oscillation, one can observe beating of
the oscillations at roughly the electron cyclotron frequency $\omega_C$. This
indicates that at least two wave modes are initially excited with a frequency
difference of $\omega_C$. After about $50\omega_{pe}^{-1}$ the wave packet,
which was 
originally located at the injection location, starts to expand into the lower
density region. Here the phase velocity $v_{\text{ph}}$ gradually descreases
from almost infinity and the lines of equal $E_y$ in the $x$--$t$ diagram start
turning from near horizontal to a finite $v_{\text{ph}}$. After about
$110\omega_{pe}^{-1}$ it becomes obvious that the group velocity $v_{\text{g}}$
is beginning to increase from near zero to a finite velocity and the wave packet
starts moving into regions of lower density. As the wave packet moves into a
lower and lower density environment, the process continues as the phase velocity
decreases and the group velocity increases. At a time of about
$500\omega_{pe}^{-1}$ the phase and group velocity have almost equalised to
$v_{\text{ph}} = v_{\text{g}} = c$, where $c$ is the speed of light. The
originally immobile wave packet has turned into an electromagnetic wave and is
allowed to freely escape.

The dynamics of the electromagnetic wave packet is almost completely decoupled
from the electron beam dynamics. While the electron beam imposes a strong $j_y$
current initially, which provides the initial conditions for the electromagnetic
field, the beam subsequently travels with a constant velocity of $0.2c$, as
prescribed by the initial conditions, into the low density region. The beam only
weakly interacts with the electromagnetic field when the wave packet starts
overtaking the beam. At around $t=200\omega_{pe}^{-1}$ the front of the wave
packet reaches the electron beam and a slight disturbance in the field can be
observed. At $t=450\omega_{pe}^{-1}$ the wave packet has almost overtaken the
beam completely. One can observe a slight enhancement of the electric field at
the beam location in the trailing part of the wave packet, but this does not
modify the wave packet as it keeps moving towards the right. One can, however,
observe a wake of the beam in the perpendicular electric field $E_y$. This wake 
does not seem to travel freely but is tied to the location of the beam and is
not converted into a free electromagnetic radiation during the course of the
simulation. 

The lower panel of figure \ref{FigEyRefPlateau} shows the $E_y$ electric field
for the simulation in which the electron beam is injected into a density
plateau. Again, one can observe an electric field oscillation at electron plasma
frequency, beating with the cyclotron frequency. The wave packet remains much
more localised to the injection region, as compared to the previous run. At
$t=100\omega_{pe}^{-1}$ the packet expands towards the right past
$x=30c\omega_{pe}^{-1}$ into the region where the density ramp starts. In the
density ramp the phase velocity can also be observed to decrease with increasing
$x$. Due to this process waves escape towards the right and finally end up
travelling with a phase speed of almost $c$ in the lower density region. In
contrast to the previous run the bulk of the wave packet remains stationary at
the injection location and only slowly loses its energy to the escaping
electromagnetic radiation. This, in turn, means that the radiation produced in
this manner is of lower 
intensity but lasts over a longer time span.

\subsection{Influence of Beam Distribution Function}

\begin{figure}
\begin{center}
\includegraphics[width=7cm]{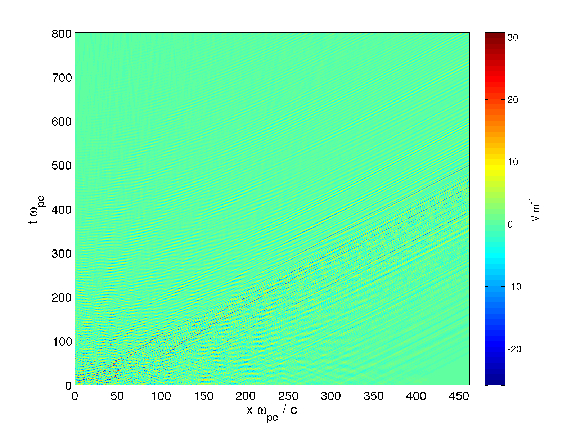}
\end{center}
\caption{The electric field component $E_y$ as a function of space and time. The
electron beam has a ring distribution.}
\label{FigEyRing}
\end{figure}

In figure \ref{FigEyRing} we plot the results for the run in which the electron
beam has a ring distribution. Here the electrons are injected into the density
ramp, just as described in the first run, and all other simulation parameters
are identical. It is evident from the plot that the initial conditions do not
excite the electromagnetic field oscillations observed in the previous two
simulations. It should be noted that the colour scale in figure \ref{FigEyRing}
has been reduced by more than a factor of 10 with respect to figure
\ref{FigEyRefPlateau}. The electric field present in the simulation can, almost
completely, be attributed to numerical noise present in the simulation. The
reason for this can be found in the fact that the initial beam distribution does
not impose a $j_y$ current on the system and, hence, no electromagnetic wave
modes are excited. Even the electromagnetic fields that the beam left in its
wake in the previous two runs are not observed when the distribution is assumed
to rotaionally symmetric in the $v_y$--$v_z$ plane. This indicates that even the
wake $E_y$ fields are a result of the particular non centered distribution
function. When the electron beam has a ring distribution function there was no
indication of an instability which could generate electromagnetic waves during
the course of the simulation,  at least by the considered simulation end-time. 

\begin{figure}
\begin{center}
\includegraphics[width=7cm]{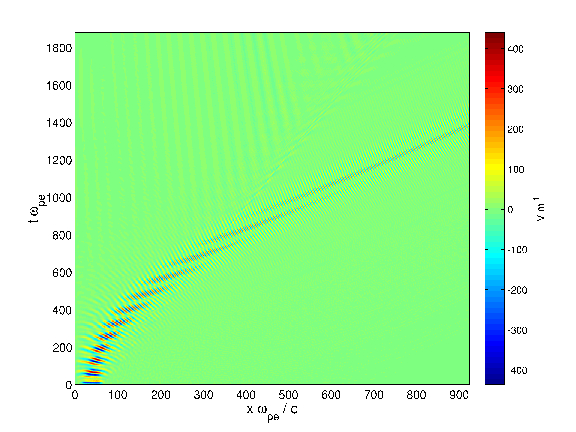}\\
\includegraphics[width=7cm]{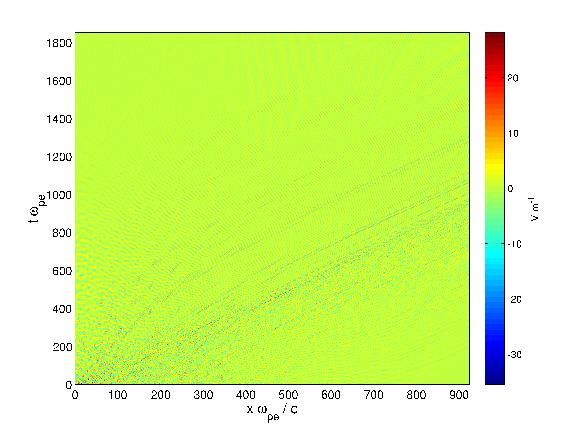}
\end{center}
\caption{The electric field component $E_y$ as a function of space and time for
the two long runs. The run with the reference parameters is shown in the top
panel, ring distributed electron beam is in the bottom panel.}
\label{FigEyLong}
\end{figure}

In order to extend the search of electromagnetic radiation by electron cyclotron
instability we performed two long runs in which both the size of the simulation
box and the simulation time was doubled. This also means that the density
gradient and the electron beam length was stretched by the same factor. Runs
were performed for the reference scenario and for the ring distribution. The
results are presented in figure \ref{FigEyLong}. The plots of the electric field
component $E_y$ for these two runs confirm all the observations of the shorter
runs. In the reference scenario the wave packet is excited by the initial
conditions and accelerates as is moves into the lower density regions. Finally,
the wave packet turns into a purely electromagnetic traveling wave and is free
to escape the system. The ring distributed electron beam, on the other hand,
does not appear to interact with the transverse electromagnetic field in any way
during the course of the simulation. This includes the apparent wake field
observed 
in the reference case, which is not present with the ring distributed beam. 

\begin{figure}
\begin{center}
\includegraphics[width=3.5cm]{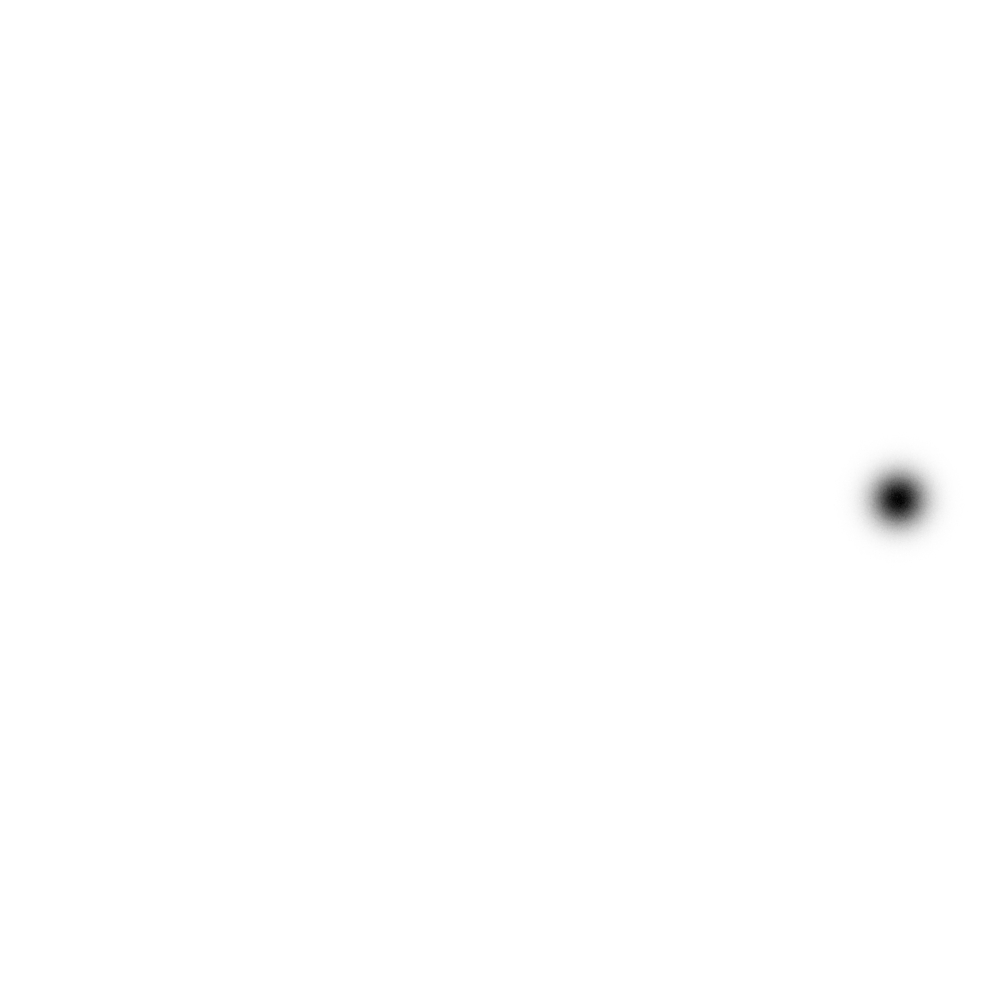}
\includegraphics[width=3.5cm]{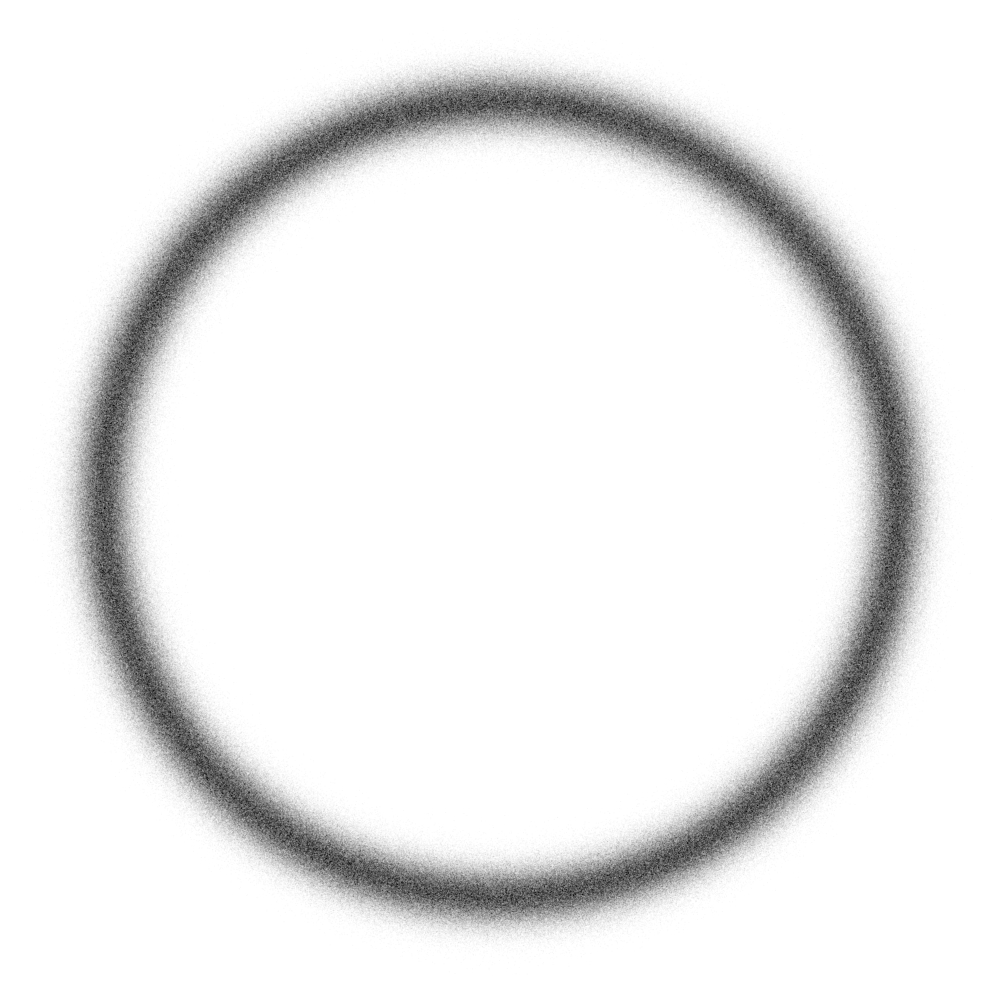}\\
\includegraphics[width=3.5cm]{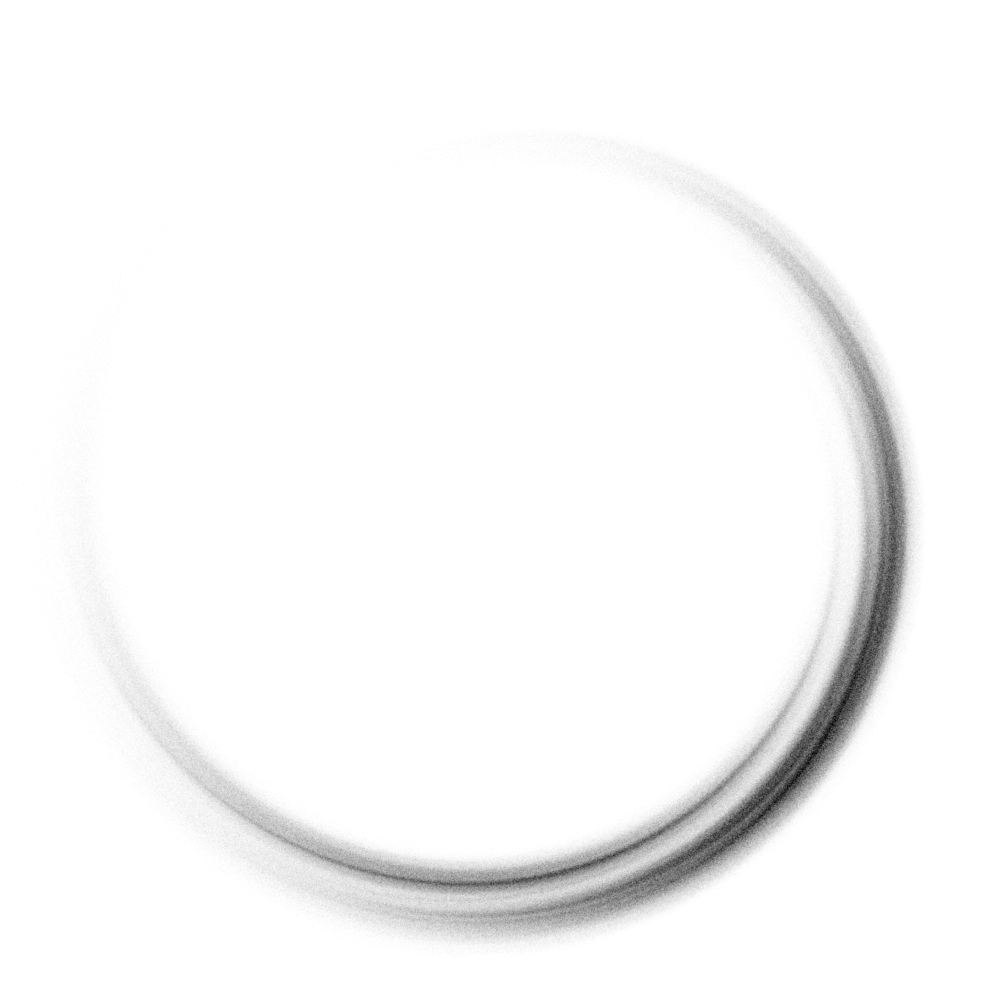}
\includegraphics[width=3.5cm]{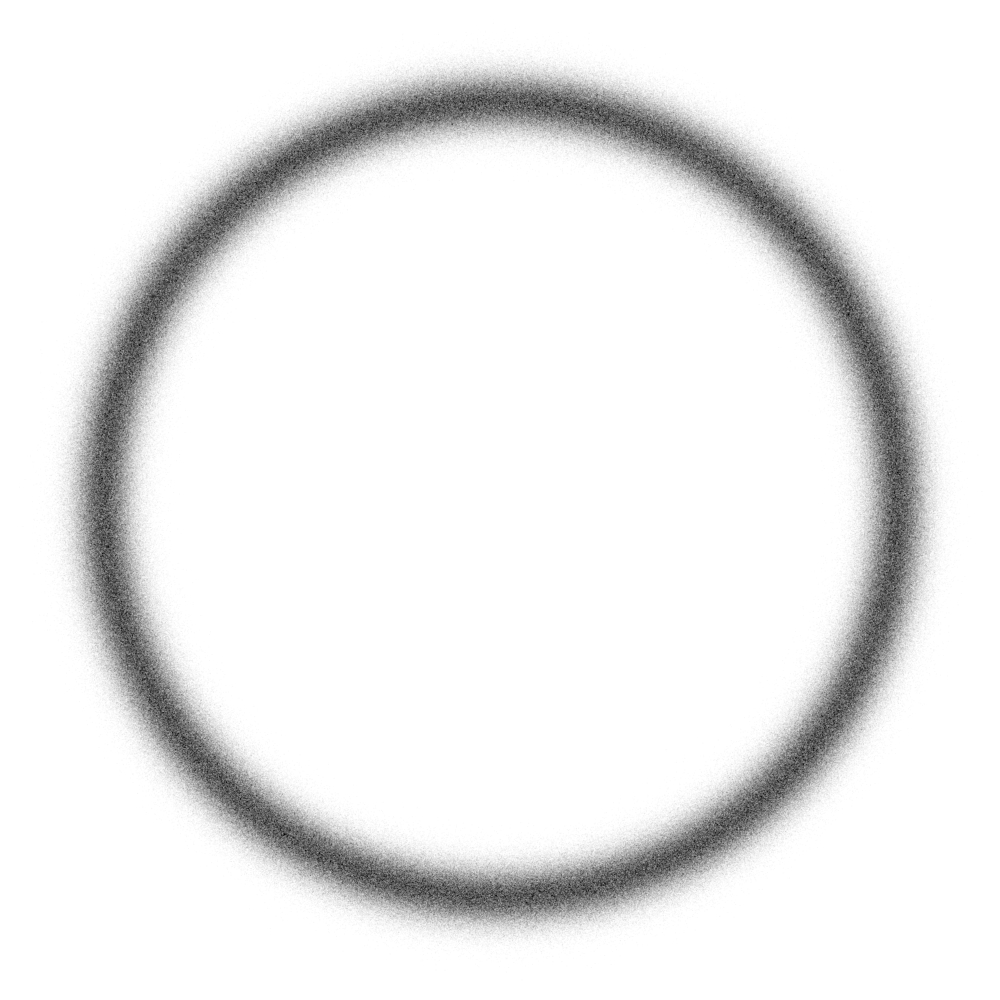}\\
\includegraphics[width=3.5cm]{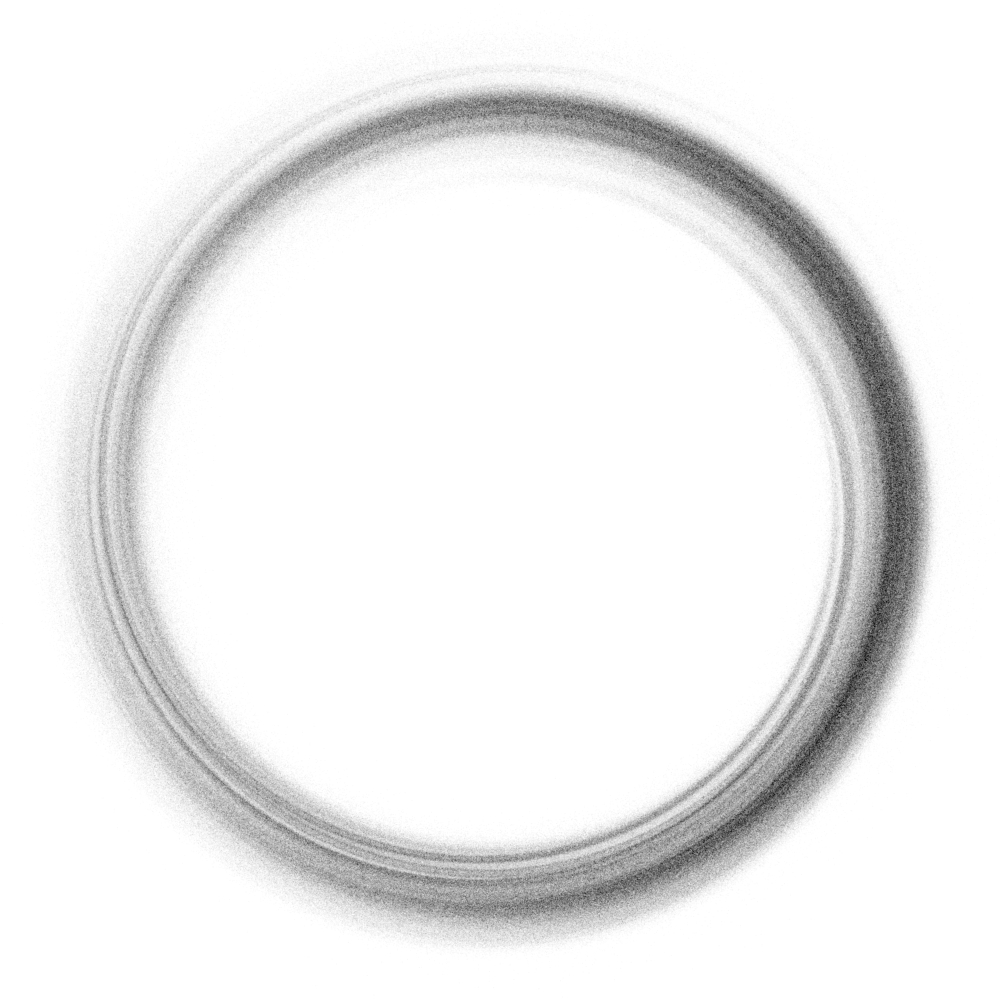}
\includegraphics[width=3.5cm]{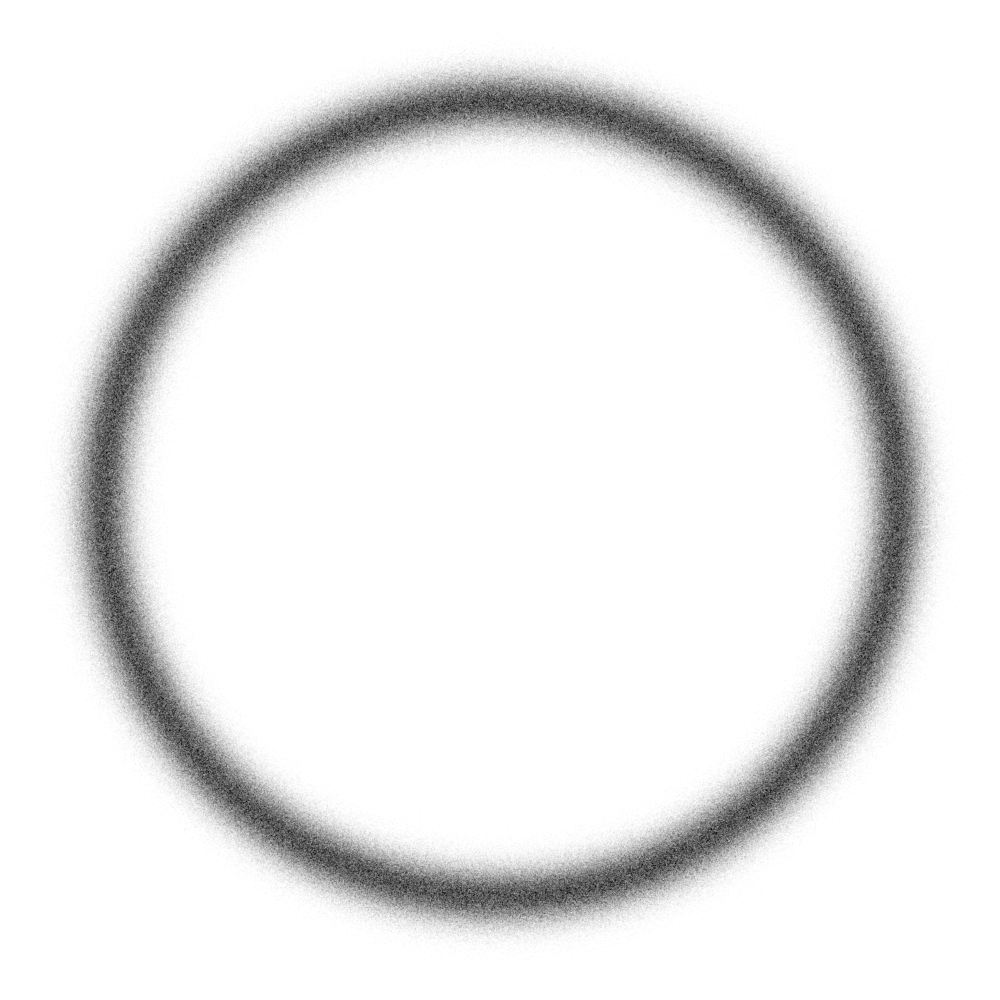}
\end{center}
\caption{The $v_y$--$v_z$ phase space densities for the long reference run (left
column) and the ring distributed beam (right column) at $t=0$ (top),
$t=900\omega_{pe}^{-1}$ (middle) and at $t=1800\omega_{pe}^{-1}$ (bottom).}
\label{FigPhasePyPz}
\end{figure}

In order to show that, in the reference run, the non-gyrotropic nature of the
distribution function is still present, while the beam travels down the density
ramp and may be responsible for the transverse electric fields, figure
\ref{FigPhasePyPz} shows the $v_y$--$v_z$ phase space densities for the two long
simulation runs. In the reference scenario the distribution function starts off
as a Maxwellian peak located at positive $v_y$ but with zero $v_z$. This peak
gyrates in phase with the cyclotron frequency. Because of fluctations in the
electromagnetic field, the strongly localised peak is gradually drawn out over a
range of phase angles. At $t=900\omega_{pe}^{-1}$ the distribution already
spreads out over a large  phase angle interval spanning more than $3\pi/2$. The
centre of the phase angle distribution is, however, still clearly visible and
the distribution resembles a crescent shape in $v_y$--$v_z$ plane. Even towards the
end of the simulation, at $t=1800\omega_{pe}^{-1}$ when the phase angles are
spread out over full 
$2\pi$, is the centre of the phase angle distribution visible. This means that,
throughout the simulation, the perpendicular current imposed by the electron
beam from the initial conditions is always present and it is this perpendicular
current which is the cause of the perpendicular electromagnetic field in the
simulations. The ring distributed electron beam is shown in the right column of
figure \ref{FigPhasePyPz}. This distribution does not carry a perpendicular
current and can, therefore, not directly excite perpendicular electromagnetic
fields. During the course of the simulation we do not observe any deviation of
the $v_y$--$v_z$ phase space distribution from the initial ring distribution.

\subsection{Analysis of Wave Modes}

\begin{figure}
\begin{center}
\includegraphics[width=7cm]{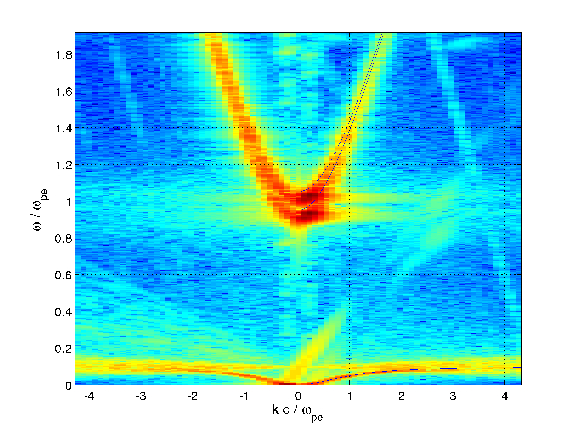}\\
\includegraphics[width=7cm]{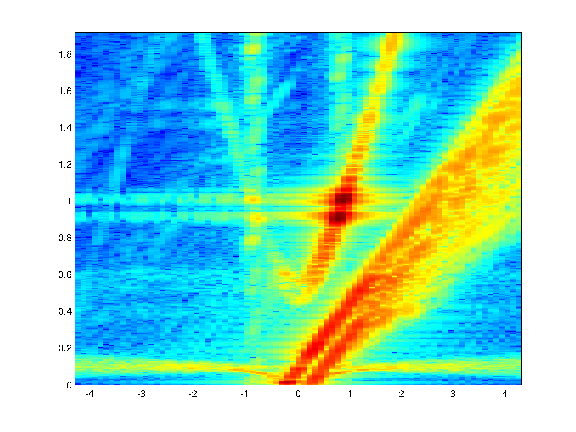}\\
\includegraphics[width=7cm]{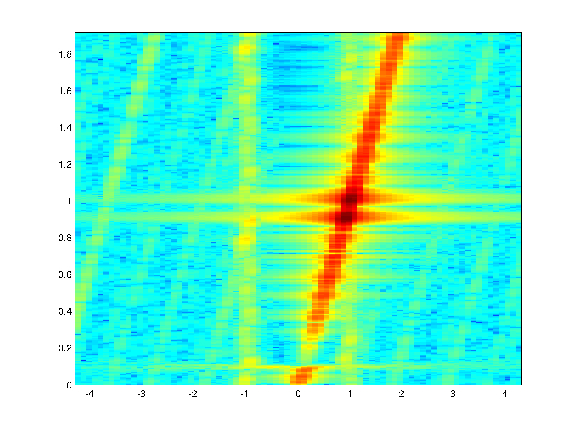}
\end{center}
\caption{Two-dimensional Fourier transform of the electric field component
$E_y$. The top panel shows the transorm at the injection region, the middle
panel near the centre of the simulation box and the lower panel at the right
edge of the box. The colour scale is logarithmic.}
\label{FigFourierRef}
\end{figure}

In order to identify the wave modes excited by the electron beam in the
reference runs and to identify the mechanism by which the initially resting wave
packet is transformed into a freely escaping electromagnetic wave, we have
performed two dimensional Fourier analyses of the electric field component $E_y$
in various regions of the simulation. The Fourier transforms were carried out
over rectangular strips in the $x$--$t$ plane with a with of $\Delta
x=50 c/\omega_{pe}$ and a length spanning the whole duration of the simulation.
Three different strips centered at $x=25 c/\omega_{pe}$, $x=250c/\omega_{pe}$,
and $x=425 c/\omega_{pe}$ were chosen to represent the injection region, the
intermediate region, and the free propagation region respectively. The result of
the Fourier transforms are shown in Figure \ref{FigFourierRef} using a
logarithmic colour scale. The advantage of a logarithmic scale is that one can
not only identify the location of the dominant modes but also, due to the
presence of numerical noise in the system, the dispersion curves on which the
maxima lie become clearly visible.

In the top panel of Figure \ref{FigFourierRef} one can see that modes are
excited in the injection region with two slightly different frequencies but
identical wave number. The superposition of these two wave modes with similar
frequency results in the beating of the electric field oscillations seen in Figure
\ref{FigEyRefPlateau}. The wave number corresponds to the length of the beam in
space and is therefore determined by the size of the initial disturbance.
Additional simulations with varying beam lengths (not shown) confirm this
notion. The maxima in the Fourier plot lie at the intersection of $k_b=2\pi/L_b$
and the dispersion curves of the L and R modes. These dispersion curves have
been calculated using the local plasma parameters, excluding any contribution  of
the electron beam, and are plotted in the top panel of Figure \ref{FigFourierRef}
as solid black curves. Very good agreement between the simulation and  analytical 
dispersion curves of the L and R modes are
found, supporting the idea that the electron beam does not influence the
propagation of the electromagnetic waves.

Because the wave modes are located near, but not quite on the minimum of the
dispersion curve the waves have a small but finite group velocity which causes
the wave packet to gradually move towards the lower density region. As the
density decreases, the dispersion curves of the L and R modes change, with the
minimum moving towards smaller frequencies. This can be seen in the middle panel
of Figure \ref{FigFourierRef}  which shows the Fourier transformed field in the
centre of the simulation domain. The frequencies of the wave modes do not appear
to change as the wave moves along the density gradient. This may be understood
in terms of the Huygens principle. As the density descreases, the dispersion relation
dictates a higher wave number, i.e. a shorter wavelength, for the same
frequency waves. This shift in wave number causes the phase velocity of the
waves to decrease. At the same time, the group velocity increases as the slope of
the dispersion curve increases near the wave modes. 

As the wave travels further down the density slope (bottom panel of Figure
\ref{FigFourierRef}) into a region of almost vanishing background density, the
dispersion relation changes into almost a linear curve corresponding to free
propagation at speed $c$. Both frequencies are still present and have not changed
noticably from the initial values but the wave number has increased further and
is now in agreement with a freely propagating electromagnetic wave. 

\section{Summary and Discussion \label{SecSummary}}

We performed extensive particle-in-cell simulations of gyrotropic and
non-gyrotropic fast electron beams in a background plasma with a decreasing
density profile. The constant magnetic field was aligned with the 1d simulation
domain. We investigated the influence of the density gradient at the injection
point and of the shape of the distribution function in the transverse plane.
Shifted Maxwellian distributions for the hot electrons with a velocity shift in
the parallel and perpendicular direction with respect to the magnetic field
resulted in a strong perpendicular current and caused electromagnetic wave modes
from the onset of the simulation. When the beam was injected into a density
gradient these waves started to accelerate and gradually turned into free
traveling electromagnetic radio waves. If, on the other hand, the electrons were
injected at a density plateau the oscillations did not travel as a wave packet
but only slowly leaked out where the tails of the localised packet reached into
regions with a density gradient.

The results presented in this paper indicate that the electromagnetic waves
observed in the simulation are purely due to the initial conditions imposed on
the system. We could not observe any instability which could be the cause of
electromagnetic radiation,  at least by the considered simulation end-time. 
In the case of an electron beam injected into a
density plateau the electromagnetic waves remained localised and there was no
further interaction of the electron beam with the background plasma visible in
the perpendicular electromagnetic fields. The electron beam is subject to the
electrostatic instabilities leading to localised oscillations in the
longitudinal electric field. Due to the 1d nature of the simulations, these
electrostatic modes cannot convert into electromagnetic modes via non-linear
wave-wave interactions. At first sight, our present results seem to contradict 
the previous investigations
suggesting a novel mechanism based on the electron cyclotron maser instability
\cite{2012PhPl...19k0702P,2012PhPl...19k2903P}. 
The latter two papers already state in their "Note added in proof" that 
whilst in their presented runs electron cyclotron maser instability
condition is met, probably the growth rate is too small for the instability
to develop by the simulation end-time.
The present results indeed confirm that the role of electron cyclotron maser instability
is insignificant and the electromagnetic emission is generated by the
transverse initial current due to the non-gyrotropic electron beam.

We stress that the transverse electromagnetic waves which
are excited in the simulation are purely due to the initial perpendicular
current imposed on the system. To support this notion we performed additional
simulations where the shifted Maxwellian was replaced by a ring distribution.
Because of symmetry, the ring distribution only carried a parallel current and
could not, therefore, excite any transverse electromagnetic waves. 

Analysis of the wave modes that were observed in the reference run showed that
the initial tranverse current excited both L and R modes with a wavenumber that
is governed by the length of the electron beam. Because the frequency of the L
and R modes differ by roughly the cyclotron frequency, a beating of the wave
packet is observed. When the initial wave packet is excited in the density
gradient the front of the packet oscillates at a slightly lower frequency than
its tail due to the change in the frequencies of the L and R modes with density.
This eventually causes the wave packet to travel down the density ramp. As the
wave packet moves into regions of lower density the phase velocity decreases
and, at the same time, the group velocity increases until, at low background
densities, the wave travels freely in the form of a vacuum electromagnetic wave.

We would like to close with the discussion of relation of the present results to the
observational aspects of solar type III radio bursts.
As mentioned in the Introduction section, the actual mechanism that generates the
radio bursts is an active line of investigation with several mechanisms put forward 
\cite{1958SvA.....2..653G,mmcp2005,1992SoPh..139..147R,2012ApJ...755...45M,2005PhPl...12e2315C,
2011PhPl...18e2903T,2012PhPl...19k0702P,2012PhPl...19k2903P}.
The key distinction of the non-gyrotropic beam mechanism \cite{2011PhPl...18e2903T} from others is 
that it does not need Langmuir waves to produce the radio emission.
As shown clearly in Refs.\cite{2012PhPl...19k0702P,2012PhPl...19k2903P}
the non-gyrotropic (i.e. non-zero pitch angle with respect to background magnetic field) electron beam perpendicular
velocity component is responsible for the EM emission generation, while
the parallel velocity component generates electrostatic Langmuir waves which also leave 
an imprint (which we refer to as beam wake) on the EM emission.
Our previous works \cite{2012PhPl...19k0702P,2012PhPl...19k2903P} made it clear that the beam
perpendicular velocity component and density gradient are needed
to generate the EM emission. However, it was not clear whether ECM instability has a dominant role or not.
Here we have shown that it is the existence of the non-zero perpendicular electron current, injected on the
density gradient are generating the EM emission and that the gradient causes wave power 
drift towards larger $k$, producing the escaping EM radiation (see Fig.~5). 
Here is how we propose the non-gyrotropic beam mechanism is realised in the solar corona:
First of all, there is an indication that solar flare-generated beam is 
not directly aligned with the ambient magnetic field, i.e. has a finite pitch angle.
In addition to the evidence discussed in Ref.\cite{2012PhPl...19k2903P}, recently, similar results
are found using the STEREO Spacecraft and the Nancay Radio Heliograph \cite{pk}.
The beam parallel velocity component generates Langmuir waves by the bump-on-tail instability, which
are detected in-situ, but have no bearing to the EM emission generation in our mechanism.
The beam perpendicular velocity component induces a non-zero transverse current, which 
then generates the EM emission in ways described above.
We should stress that, as can be seen from the left column of Fig.~4, the electron transverse
velocity phase-space becomes ring-like very quickly, by $t=1800 \omega_{pe}^{-1}$, thus
shutting off the EM emission generation, as the ring gives zero net transverse current.
This means that our mechanism predicts that typical coronal type III burst
durations of, say, 1 s, is {\it related to the time-scale of the electron beam continuous-in-time
re-injection}. In the plasma emission mechanism, the inverse of the growth rate of the bump-on-tail instability, i.e.
quali-linear relaxation time, which would be good estimate for EM emission time  
is $\tau=1/\gamma=n_{background}/(n_{beam} \omega_{pe})$. 
With $\omega_{pe}^{-1}$ being, say, $10^{-8}$ s, in oder to get 1 s duration EM emission, one would need 
$n_{beam}/b_{background}=10^{-8}$ -- a very dilute beam indeed. The latter may be plausible, but in the
corona we cannot measure beam density directly and one has to make do with postulating  the range $n_{beam}/b_{background}=10^{-5}-10^{-8}$.
In the non-gyrotropic beam mechanism, however, because $t=1800 \omega_{pe}^{-1} \approx 10^{-5}$s, the duration of emission is
then related to the beam re-injection time, so it serves is a direct diagnostic of the beam injection time-dynamics.
Of course, one has to remember that our model is 1.5D and  when more that one spatial dimension is allowed, then of course
the plasma emission mechanism becomes allowed (as discussed in Ref.\cite{2011PhPl...18e2903T} 1D switches off the plasma emission).
Thus, the interplay between plasma emission and non-gyrotropic beam mechanisms depends on 
the curvature of the background magnetic field: in the field that is purely radial (essentially 1D in space), our mechanism would be
dominant, whereas if the transverse curvature is allowed, then both mechanisms can be in action -- this depends on the pitch angle
i.e. at what angle the electron beam is injected. The key message here is that the non-gyrotropic beam mechanism
divorces the radio emission burst duration from plasma kinetics (which is too short-scale in time) and relates it to
the electron beam re-injection time scale, which may be an important diagnostic of the beam generation, i.e.
whether it is driven by magnetic reconnection process or the electron acceleration by the dispersive Alfven waves \cite{dawp}. 
 
\begin{acknowledgments}
This work is funded by Leverhulme Trust research grant RPG-311.
Computational facilities used are that of Astronomy Unit, 
Queen Mary University of London and STFC-funded UKMHD consortium at St Andrews
and Warwick
Universities. DT is financially supported by STFC consolidated grant
ST/J001546/1
and HEFCE-funded South East Physics Network (SEPNET).
\end{acknowledgments}

%\bibliographystyle{myaip}       % APS-like style for physics
%\bibliography{references}   % name your BibTeX data base

\end{document}